\documentclass[sigplan,10pt]{acmart}
\acmConference[Preprint]{Preprint}{2026}{}
\acmYear{2026}
\acmISBN{}
\acmDOI{}
\startPage{1}
\setcopyright{none}
\usepackage{subcaption}
\usepackage{colortbl}
\usepackage{array}
\usepackage{tabularx}
\usepackage{pifont}
\usepackage{booktabs}
\usepackage{multirow}
\usepackage[table]{xcolor}
\usepackage{hyperref}

\graphicspath{{figures/}}

\setcitestyle{numbers,sort&compress}

\renewcommand\footnotetextcopyrightpermission[1]{}

\newcommand{\cmark}{\ding{51}}
\newcommand{\xmark}{\ding{55}}

\author{Zhixiang Wei}
\affiliation{%
\institution{Shanghai Jiao Tong University}
\city{Shanghai}
\country{China}
}
\email{tonywei_sjtu@sjtu.edu.cn}

\author{Yun Wang}
\affiliation{%
\institution{Shanghai Jiao Tong University}
\city{Shanghai}
\country{China}
}
\email{yunwang94@sjtu.edu.cn}

\author{James Yen}
\affiliation{%
\institution{Shanghai Jiao Tong University}
\city{Shanghai}
\country{China}
}
\email{jamesyen2202002@gmail.com}

\author{Mingyuan Xia}
\affiliation{%
\institution{UltraRISC}
\city{Shanghai}
\country{China}
}
\email{xiamy@ultrarisc.com}

\author{Zhengwei Qi}
\affiliation{%
\institution{Shanghai Jiao Tong University}
\city{Shanghai}
\country{China}
}
\email{qizhwei@sjtu.edu.cn}

\begin{document}

\title{HBM Is Not All You Need: Efficient Disaggregated LLM Serving across Memory-heterogeneous Accelerators}

\renewcommand{\shortauthors}{Wei et al.}

\begin{abstract}
LLM inference comprises a compute-bound prefill phase and a memory-bound decode
phase, and recent systems disaggregate them onto separate hardware. Yet today's
datacenter GPUs rely on costly HBM whose bandwidth sits almost entirely idle
during prefill. LLM serving across memory-heterogeneous accelerators (MemHA) pairs GDDR-based accelerators for prefill with HBM-based GPUs for decode, promising lower cost without sacrificing performance. Pushed to its most economical form, MemHA serving is inherently cross-vendor, since the best-suited chip for each phase may come from a different vendor. This breaks two assumptions that
single-vendor disaggregation takes for granted---a KV format both ends consume
natively, and a shared software stack. We present \textbf{HMA-Serve},
a MemHA-centric disaggregated serving system pairing GDDR-based accelerators for prefill with
HBM-based GPUs for decode efficiently. HMA-Serve achieves this through (1)
phase-wise quantization, applying vendor-native low precision for high-throughput prefill while keeping decode in high-precision BF16,
(2) a compute-transfer pipeline that overlaps each layer's KV cache transfer
with later-layer prefill to reduce time-to-first-token (TTFT), and (3)
deferred dequantization, shipping raw quantized bytes and reconstructing
them lazily on the decode GPU to reduce network bandwidth and HBM usage. Across four Qwen3 models (4B--32B) and three production traces,
HMA-Serve delivers up to $3.2\times$ higher goodput than state-of-the-art memory-homogeneous methods and $4.8\times$ higher goodput-per-dollar, with no measurable loss
on generation-quality benchmarks.
\end{abstract}

\maketitle

\section{Introduction}

Large Language Models (LLMs) are now deployed across chatbots, code assistants, and
agentic applications, but serving them is extremely expensive. One early estimate
placed the daily cost of running ChatGPT at roughly \$700{,}000 in early 2023,
requiring on the order of 30{,}000 NVIDIA A100 GPUs, and demand has only grown
since. As of early 2026, NVIDIA's flagship accelerators remain sold out with a
multi-million-unit backlog, roughly two-thirds of which is expected to serve
inference. Driving down the hardware cost of LLM serving is therefore a
first-order systems problem.

LLM inference runs in two phases with opposing resource profiles. The
\emph{prefill} phase processes the entire prompt in parallel to produce the KV
cache and the first token, and is compute-bound. The \emph{decode} phase generates
tokens sequentially, each depending on the KV state of all preceding tokens, and is
memory-bound, sweeping the KV cache and weights once per token. Because each phase
exercises only a subset of a GPU's resources, prior work improves utilization in
two ways. \emph{Prefill-decode colocation} batches prefill and decode of different requests to
share weights, but suffers from prefill--decode interference and tail latency. \emph{Prefill-decode disaggregation} runs the phases on separate instances connected by a
KV cache transfer, removing interference and matching each phase to suitable
hardware at modest transfer cost.

\textbf{Despite these software optimizations, hardware efficiency remains limited by a
mismatch between prefill and costly HBM.} HBM has become the dominant cost component
of modern accelerators---18\% of A100 manufacturing cost, rising to 45\% on
B200---yet prefill's high arithmetic intensity leaves this premium bandwidth almost
entirely idle. We measure an A100 wasting over 97\% of its HBM bandwidth during
4K-token prefill. This motivates \emph{Memory-Heterogeneous Accelerator} (MemHA)
serving: GDDR-based accelerators, such as Tenstorrent's chips, handle prefill,
while HBM-based GPUs handle decode. This approach is highly promising in principle. For example, a state-of-the-art GDDR-based accelerator, the Tenstorrent Blackhole p150, provides 664 TFLOPS of Blocked FP8 (BFP8) compute at roughly \$1,300, an order of magnitude cheaper than the A100 while offering comparable or higher low-precision throughput. Because cost-efficient GDDR accelerators have only
recently shipped and share no software stack with NVIDIA, the most economical MemHA
pairing is inherently \emph{cross-vendor}, and little is known about whether and
how it pays off.

Making cross-vendor MemHA serving efficient raises two systematic challenges.
\textbf{First, the KV cache sits on the critical path of every request and is large
enough to dominate latency unless its transfer is hidden.} Qwen3-32B at 8K context
produces gigabytes of KV state; even over a 100~Gbps RDMA fabric, a monolithic
post-prefill transfer takes hundreds of milliseconds---comparable to prefill
itself---and a naive schedule charges all of it to time-to-first-token (TTFT). The
cross-vendor path is also longer than a network hop. The cache must first cross a
device-to-host DMA before reaching the fabric. \textbf{Second, the KV cache a GDDR
prefiller produces does not match what HBM decode kernels expect, in number format
or memory layout.} Tenstorrent stores BFP8 in a $32{\times}32$ tiled layout, while
NVIDIA decode kernels read BF16 in the paged row-major layout FlashAttention~\cite{flashattention}
consumes. Both naive bridges are bad: dequantizing on the producer doubles wire
traffic, while a standalone consumer pass adds a full-tensor HBM read that competes
with attention.

We present \textbf{HMA-Serve}, which turns the precision asymmetry into a
performance lever and keeps transfer off the critical path through three
coordinated mechanisms. \emph{Phase-wise quantization} runs prefill in
Tenstorrent's vendor-native BFP8 and decode in
BF16 to maximize throughput while minimizing accuracy loss. \emph{Compute-transfer pipelining} exposes
per-layer completion events from the prefill runtime and overlaps each layer's KV
egress---device-to-host push plus RDMA---with the prefill of later layers.
\emph{Deferred dequantization} ships raw quantized bytes verbatim, halving wire
traffic, and reconstructs them lazily inside a fused decode-side kernel; because the
reconstruction is integer bit manipulation, it runs on the GPU's integer ALU rather
than the tensor cores decode saturates, isolating its cost from decode at the
hardware level instead of competing for the same compute.

We implement HMA-Serve on a deployed cluster of four Tenstorrent Blackhole p150b
accelerators for prefill and an NVIDIA A100~80\,GB for decode over a 100~Gbps RDMA
network, and evaluate four Qwen3 models (4B--32B) under three production traces.
This work contributes:
\begin{itemize}
\item \textbf{HMA-Serve}, to our knowledge the first MemHA serving system on real
silicon pairing GDDR-based prefill with HBM-based decode across a commodity RDMA
fabric.
\item \textbf{Phase-wise quantization}: vendor-native low precision for
compute-bound prefill, BF16 for precision-sensitive decode.
\item A \textbf{Compute-transfer pipeline} that hides cross-vendor KV egress behind
later-layer prefill.
\item \textbf{Deferred dequantization}: shipping raw quantized bytes and
reconstructing them lazily in a fused decode-side kernel.
\item An evaluation across four LLMs and three workloads showing up to $3.2\times$
higher goodput than state-of-the-art memory-homogeneous serving systems and $4.8\times$ higher
goodput-per-dollar, with no measurable quality loss.
\end{itemize}

\section{Background and Motivation}

\autoref{tab:design-space} positions HMA-Serve against representative serving
systems along five axes: off-chip memory architecture, heterogeneous hardware
across phases, disaggregation, phase-wise precision, and real-silicon
demonstration.

\begin{table}[t]
\centering
\footnotesize
\setlength{\tabcolsep}{3pt}
\renewcommand{\arraystretch}{1.15}
\begin{tabular}{@{}l@{\hspace{5pt}}ccccc ccc@{}}
\toprule
 & \textbf{Non-HBM} & \textbf{Het.} & \textbf{Dis.} & \textbf{PQ} & \textbf{Real} & \textbf{Lat.} & \textbf{Thru.} & \textbf{Cost} \\
\midrule
Orca~\cite{orca}        & \xmark & \xmark & \xmark & \xmark & \cmark & Var.  & Low  & High \\
Sarathi~\cite{sarathi}  & \xmark & \xmark & \xmark & \xmark & \cmark & Var.  & High & High \\
Groq~\cite{groq} & \cmark & \xmark & ?      & \xmark & \cmark & Low   & ?    & ?    \\
DistServe~\cite{distserve} & \xmark & \xmark & \cmark & \xmark & \cmark & Low & High & Med  \\
Splitwise~\cite{splitwise} & \xmark & \cmark & \cmark & \xmark & \cmark & Low & High & Med  \\
Mix-Quant~\cite{mixquant}  & \xmark & \xmark & \cmark & \cmark & \cmark & Low & High & Med  \\
SPAD~\cite{spad}        & \cmark & \cmark & \cmark & \xmark & \textit{sim} & Low & High & Low  \\
\rowcolor{gray!12}
\textbf{HMA-Serve}     & \cmark & \cmark & \cmark & \cmark & \cmark & Low   & High & Low  \\
\bottomrule
\end{tabular}
\caption{\textbf{LLM serving design space.} \textbf{Non-HBM}: a phase chip uses
GDDR/SRAM; \textbf{Het.}: heterogeneous hardware across phases; \textbf{Dis.}:
disaggregated scheduling; \textbf{PQ}: phase-wise quantization; \textbf{Real}:
real silicon; \textbf{Cost}: cost per goodput.}
\label{tab:design-space}
\end{table}

\subsection{LLM Serving Hardware}
Accelerators used for LLM serving fall into three families distinguished by their
off-chip memory. \textbf{HBM-based} accelerators (NVIDIA A100/H100/H200/B200, AMD
MI300X, Google TPU) integrate stacked-die DRAM on package, delivering
2--8\,TB/s across 80--192\,GB. This bandwidth is essential for memory-bound decode,
where every token sweeps the whole KV cache and weights, but HBM is structurally
expensive: HBM3e contracts at roughly \$15/GB---about $3\times$ commodity GDDR---and
now accounts for 35--45\% of an accelerator's manufacturing cost. \textbf{GDDR-based}
accelerators (Tenstorrent Wormhole/Blackhole, NVIDIA L40/L4, the announced Rubin
CPX) solder GDDR on a standard PCB, trading roughly $3\times$ lower per-GB cost for
lower bandwidth---well-matched to compute-bound prefill, whose high arithmetic
intensity leaves extra bandwidth idle, but too bandwidth-starved to decode on their
own. \textbf{SRAM-based} accelerators (Groq LPU~\cite{groq}, Cerebras WSE-3) place all working
memory in on-die SRAM, reaching hundreds of TB/s but only tens of GB, so serving a
70B-class model demands aggressive sharding that inflates inter-chip cost.

\begin{figure}[t]
    \centering
    \includegraphics[width=\linewidth]{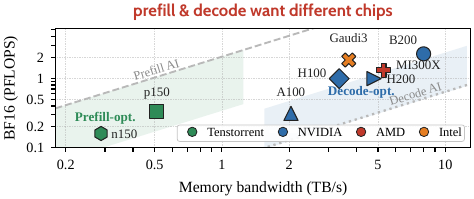}
    \caption{\textbf{Roofline of available, scalable AI chips across vendors.}
    Overlaying prefill's and decode's arithmetic-intensity rays partitions the
    landscape: GDDR parts (e.g.\ Tenstorrent p150) sit in the compute-rich,
    bandwidth-modest region ideal for prefill; HBM GPUs in the bandwidth-rich
    region ideal for decode. The most cost-efficient design pairs the two rather
    than buying one chip that does both.}
    \label{fig:roofline}
    \vspace{-1em}
\end{figure}

\begin{figure*}[t]
    \centering
    \includegraphics[width=\textwidth]{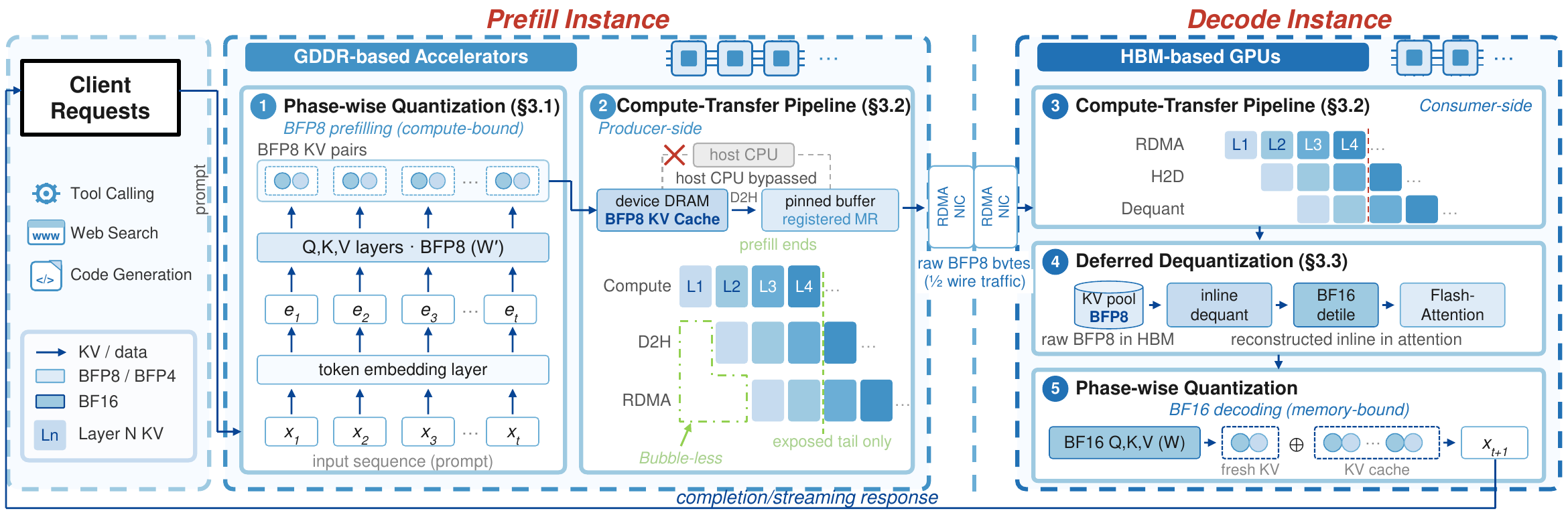}
    \caption{\textbf{Architecture of HMA-Serve.} A scheduler routes each request
    to a Tenstorrent prefill worker and an A100 decode worker over a 100~Gbps RoCE
    link. \textbf{(1)} Each phase runs at its native precision---BFP8 prefill,
    BF16 decode (\S\ref{sec:pq}). \textbf{(2)} A compute-transfer pipeline overlaps
    each layer's compute, device-to-host push, and RDMA on the producer with RDMA
    receive, host-to-device copy, and dequantization on the consumer
    (\S\ref{sec:pipe}). \textbf{(3)} The decode worker keeps the KV cache as raw
    BFP8 pages in HBM and reconstructs them to BF16 inline inside the paged
    attention kernel (\S\ref{sec:deq}). Lighter shades mark BFP8 values, darker
    shades BF16.}
    \label{fig:arch}
\end{figure*}

\subsection{Motivation}
Prefill amortizes each weight across $L$ prompt tokens, so its arithmetic intensity
scales with $L$ and becomes compute-bound past $\approx\!1$K tokens; decode reuses
each weight once per token, leaving arithmetic intensity near 1 and firmly
memory-bound. \autoref{fig:roofline} plots representative accelerators and overlays
these rays, partitioning the landscape into a compute-rich, bandwidth-modest region
where GDDR parts like Tenstorrent's p150 sit, a bandwidth-rich region where HBM GPUs
sit, and a costly upper-right region delivering both. The roofline therefore
predicts that the cost-efficient design pairs GDDR prefill with HBM (or even SRAM)
decode rather than a single chip carrying both.

\textbf{HBM bandwidth is wasted in prefill.} Running dense prefill on a single A100
across four Qwen3 sizes (4B--32B) in BF16, MFU exceeds 70\% past 4K for all four
models while MBU falls below 10\% around 1K and under 1\% at 16K. At $L=4$K the A100
leaves 96.8--97.2\% of its HBM bandwidth idle, rising above 99\% at 16K---any memory
that keeps up with prefill's modest effective bandwidth, a bar GDDR comfortably
meets, suffices.

\textbf{GDDR keeps pace in practice.} Real performance also depends on kernel
quality and dispatch overhead, so we measure end-to-end prefill latency for
Qwen3-32B on a four-card Tenstorrent mesh against a tuned single-A100 vLLM
deployment with CUDA graphs. Within Tenstorrent's \emph{trace mode}, where prefill
runs from a prebaked dispatch graph for $L\le4$K, the mesh beats the A100 by
$1.33$--$1.38\times$; at 8K, where the runtime falls back to eager per-op dispatch,
it still leads by $1.19\times$; the A100 reclaims the advantage only at $\ge\!16$K,
and there by under $1.07\times$. GDDR silicon is thus empirically competitive across
the practical prefill range, at roughly one-tenth the per-chip price.

\subsection{Challenges and Opportunities}
\textbf{Challenge 1: KV transfer adds TTFT if scheduled naively.} The entire KV
cache produced by prefill must cross the inter-accelerator interconnect before
decode begins. Qwen3-32B at 8K is $\approx\!2$\,GB in BF16 (1\,GB in BFP8); over a
100~Gbps RDMA fabric ($\approx\!10$\,GB/s effective) a monolithic transfer costs
100--200\,ms, 10--15\% of the prefill latency, charged in full to TTFT by a naive
schedule. The opportunity is structural: prefill runs layer by layer, layer
$\ell$'s KV is ready once its attention finishes, and its transfer is independent
of layers $\ell{+}1,\ldots$, so it can be pipelined behind later compute.
Overlapping transfer with compute is established practice in homogeneous systems
(Splitwise~\cite{splitwise}, TensorRT-LLM), but two properties make our setting harder: the
cross-vendor path adds a device-to-host DMA on the producer before the fabric---a
large fraction of transfer cost---and no prior system overlaps transfer across a
cross-vendor boundary, since all assume identical accelerators emitting and
consuming the same KV format.

\textbf{Challenge 2: the two sides' number formats are incompatible, and naive
bridging defeats the purpose.} This is a property of the accelerator landscape, not
of our two chips: each matrix engine reaches peak throughput only at its own native
low-precision format (NVIDIA moves FP16$\to$FP8$\to$NVFP4 across A100/H100/B200; AMD
moves FP8$\to$MXFP4 across MI300X/MI355X), and these formats rarely agree across
vendors or even generations. Any MemHA pairing therefore spans at least two native
formats, so pinning the cluster to one global precision parks at least one side off
its efficient point. Concretely, reconciling Tenstorrent's tiled BFP8 with the
paged-BF16 layout NVIDIA kernels and FlashAttention consume requires choosing
\emph{which} precision each phase runs in and \emph{where} conversion happens, and
both naive choices are bad: producer-side dequant doubles wire payload and pushes a
conversion pass onto the prefill critical path; a standalone consumer pass adds a
full KV read that competes with attention for HBM bandwidth. The opportunity is
that prefill is low-precision-tolerant while decode is precision-sensitive, and the
precision conversion is integer bit manipulation that maps onto the GPU's integer
ALU---a separate hardware unit from the tensor cores decode saturates---so the KV
can be shipped raw and reconciled lazily on the consumer without contending with
decode for compute.

\section{Design}

\autoref{fig:arch} gives the end-to-end picture. A disaggregation scheduler routes
each request to a Tenstorrent prefill worker and an A100 decode worker connected by
a 100~Gbps RoCE link, and three mechanisms turn the cross-vendor split into a
performance lever: each phase runs at its hardware's native precision
(\S\ref{sec:pq}); a compute-transfer pipeline hides KV egress behind prefill
(\S\ref{sec:pipe}); and the decoder reconstructs BFP8 pages lazily inside paged
attention (\S\ref{sec:deq}).

\subsection{Phase-wise Quantization}
\label{sec:pq}
The two phases differ in precision sensitivity in a way that aligns with their
hardware. Prefill is compute-bound and tolerant of low precision, while decode is
memory-bound and sensitive to error accumulation over long generations. HMA-Serve
therefore runs prefill in Tenstorrent's vendor-native BFP8---mixed BFP8/BFP4
weights with BFP8 activations and KV cache---and keeps decode in BF16 on the A100.
This keeps each side at its peak-efficiency point. BFP8 weights speed the
matmul-bound prefill by a measured $1.1$--$1.5\times$ over a BF16-uniform
configuration (the gain is largest at short prompts, where weight-fetch bandwidth
dominates) and halve prefill-side weight memory, freeing GDDR capacity for KV and
activations. Keeping decode in BF16, meanwhile, preserves generation
quality (\S\ref{sec:eval-accuracy}): quantizing the long, error-accumulating decode
phase is exactly what costs accuracy on the hardest reasoning and long-context
tasks.

\begin{table*}[t]
\centering
\footnotesize
\setlength{\tabcolsep}{6pt}
\begin{tabular}{@{}llcccc@{}}
\toprule
Model & Scenario & Dataset & In\,/\,Out (tok) & Prefill Setup & TTFT\,/\,TPOT SLO \\
\midrule
Qwen3-4B  & Decode-heavy        & ShareGPT  & $512\,/\,512$  & DP$=4$ & $0.1$\,s\,/\,$60$\,ms \\
Qwen3-8B  & Balanced            & ShareGPT  & $1024\,/\,384$ & DP$=4$ & $0.8$\,s\,/\,$70$\,ms \\
Qwen3-14B & Prefill-heavy       & LongBench & $2048\,/\,256$ & DP$=2$, TP$=2$ & $2$\,s\,/\,$120$\,ms \\
Qwen3-32B & Prefill-heavy$^{+}$ & arXiv     & $4096\,/\,64$  & TP$=4$ & $7$\,s\,/\,$240$\,ms \\
\bottomrule
\end{tabular}
\caption{\textbf{Evaluation setups.} Four Qwen3 sizes, each driven by a
representative open trace at a fixed input/output length spanning decode-heavy
chat to prefill-heavy summarization. \emph{Prefill} is HMA-Serve's
data-/tensor-parallel layout on the four-chip Tenstorrent mesh (every layout uses
all four chips, DP$\times$TP$=4$); decode is always one BF16 A100. Both SLOs are
$5\times$ the no-load single-request latency of the strongest baseline
(DistServe-Homo).}
\label{tab:eval-setup}
\end{table*}

\subsection{Compute-Transfer Pipeline}
\label{sec:pipe}
A monolithic post-prefill KV transfer would add 100--200\,ms to TTFT. HMA-Serve
hides it by pipelining at layer granularity across the full cross-vendor path. It
monkey-patches the otherwise-opaque prefill runtime to expose a per-layer
completion event, and as each layer's KV is produced it is \emph{gathered} into a
device-DRAM staging buffer---an operation folded into the prefill execution trace,
so it adds essentially no latency. A fast device-to-host \emph{push} kernel then
evacuates the staged bytes to pinned host memory: eight Tensix data-movement cores
write the raw BFP8 tile bytes straight into the pinned RDMA buffer over PCIe at a
measured $\approx\!6.7$\,GB/s---about $7\times$ the standard host-pulled readback
($\approx\!0.93$\,GB/s). A pool of
RDMA streams then ships each layer over the fabric while later layers are still
prefilling, so most of the cache has landed on the decode side by the time prefill
finishes; because the NIC reads the pinned buffer directly, the host CPU stays off
the path. The decode side symmetrically overlaps RDMA receive, host-to-device copy,
and dequantization. At a measured $92.3$\,Gb/s the per-request egress is only
$1.5$--$49$\,ms across our sizes---small against the $\sim\!0.9$\,s prefill---so for
a single request the overlap saves little, but it is what lets egress hide behind
the \emph{next} request under sustained load, which is where throughput is set.

\subsection{Deferred Dequantization}
\label{sec:deq}
Rather than converting BFP8$\to$BF16 on the producer (which doubles wire traffic)
or in a standalone consumer pass (which adds a full HBM read), HMA-Serve ships the
raw BFP8 bytes verbatim and keeps them as raw BFP8 pages in the decoder's HBM,
reconstructing them lazily inside the paged-attention kernel. Reconstruction fuses
four steps into a single pass: block-float decode of the BFP8 tiles (extract the
per-block shared exponent and 7-bit mantissa) to BF16; the tiled$\to$row-major
layout conversion that merges the per-chip KV shards; the post-RoPE key re-layout
that bridges Tenstorrent's interleaved head-dimension order to the half-split order
FlashAttention expects (keys only---values carry no rotary embedding); and the
first paged read. Crucially, BFP8$\to$BF16 is not floating-point arithmetic but
block-float \emph{bit manipulation}---shifts, masks, ORs, and a bitcast to assemble
the IEEE pattern---so it runs on the integer ALU that GEMM-heavy decode leaves
nearly idle, on disjoint hardware from the tensor cores. A fused kernel rebuilds a
whole request's KV in $\sim\!1.5$\,ms (versus $72$\,ms for a naive elementwise
reconstruction), so at serving rates the dequant appears as sparse, brief bursts
that overlap decode rather than a sustained load, and decode throughput stays within
$1\%$ of a dequant-free baseline even at the busiest batch that fits in HBM.

\begin{figure*}[t]
    \includegraphics[width=\linewidth]{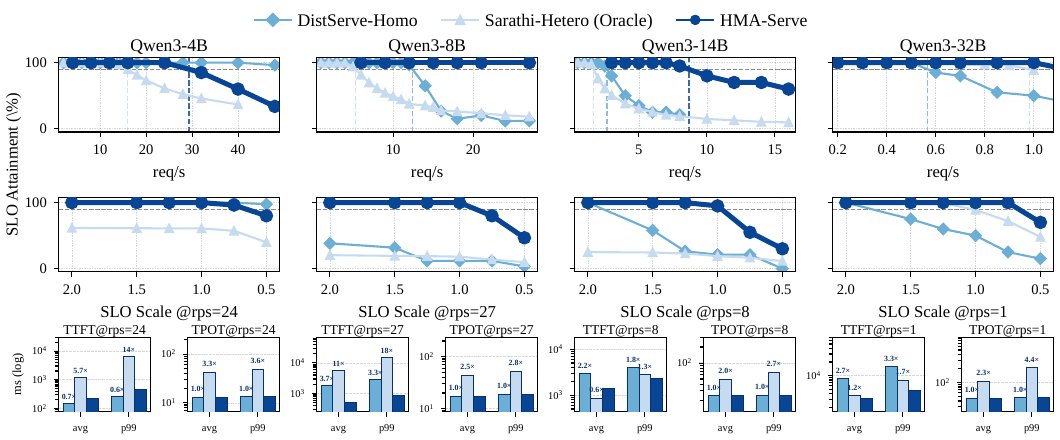}
    \caption{\textbf{HMA-Serve vs.\ the strongest disaggregation
    (DistServe-Homo) and colocation (Sarathi-Hetero) baselines---three metrics
    $\times$ four model scales}, all measured end-to-end on real silicon.
    \textbf{Top:} SLO attainment vs.\ offered rate (dashed line at $90\%$; each
    system's vertical mark is its $90\%$ crossover). \textbf{Middle:} attainment
    as the joint TTFT/TPOT SLO is relaxed (looser toward the left), at the
    reference rate. \textbf{Bottom:} TTFT and TPOT (avg and p99) at the reference
    rate, on a log scale---at each model's HMA-Serve gp@90 rate, the baselines
    are already past their SLO limit, so their elevated latency is the goodput gap
    expressed in latency.}
    \label{fig:goodput_4way}
\end{figure*}

\section{Evaluation}

\textbf{Testbed and systems.} We evaluate every system on real silicon: NVIDIA
A100-80GB on the HBM side, a four-chip Tenstorrent Blackhole p150 mesh (TT$\times$4)
on the GDDR side, over a 100~Gb RoCE fabric, all serving through
vLLM~0.19.1~\cite{vllm}. We disable prefix caching everywhere (the GDDR prefill path
keeps none) and pin the Tenstorrent core clock to its rated frequency so prefill
timing is not distorted by idle down-clocking. \textbf{HMA-Serve} (ours) runs BFP8
prefill on TT$\times$4 and BF16 decode on one A100, mapping each model's prefill
onto the mesh with the data-/tensor-parallel layout of \autoref{tab:eval-setup}. \textbf{DistServe-Homo} is
homogeneous 1P1D disaggregation across two A100s with real GPUDirect-RDMA KV
transfer (no host staging)---the strongest same-vendor baseline. We calibrate the
relative SLOs to its no-load latency.
\textbf{Sarathi-Hetero (Oracle)} applies chunked-prefill colocation on both an A100
and the TT$\times$4 mesh, routing each request to whichever of the two---at whatever
split---maximizes SLO-meeting throughput; since the split extremes recover either
alone, it upper-bounds any static colocation policy.

\textbf{Metric.} A request is \emph{served} only if it meets both its TTFT and TPOT
SLOs, each set to $5\times$ the corresponding no-load single-request latency of the
strongest baseline (DistServe-Homo)---the tightest, most defensible bar, following
the relative-SLO convention of Sarathi-Serve~\cite{sarathi}, Splitwise~\cite{splitwise}, and Mooncake~\cite{mooncake}. Our primary
metric is \emph{goodput at $90\%$ attainment} (gp@90): the highest served
throughput at which $\ge\!90\%$ of requests still meet both SLOs. We sweep four
Qwen3 sizes (4B--32B), each with a representative open workload---ShareGPT for chat,
LongBench for long-context QA, and sampled arXiv articles for summarization---under
Poisson arrivals at fixed per-regime input/output lengths spanning short-prompt/
long-generation chat to long-prompt/short-generation summarization.

\smallskip\noindent\textbf{Workloads and prefill parallelism.}
\autoref{tab:eval-setup} lists the four setups---trace, input/output length, and
the two relative SLOs---together with how HMA-Serve maps each model's prefill onto
the four-chip Tenstorrent mesh. A single short-prompt prefill underuses the mesh,
so 4B and 8B run \emph{data-parallel} as four 1-chip replicas (DP$=4$, TP$=1$);
14B needs two chips per replica for capacity (DP$=2$, TP$=2$); and the 4K-token
32B prefill fills the whole mesh as one tensor-parallel group (DP$=1$, TP$=4$).
Every configuration uses all four chips (DP$\times$TP$=4$), and decode is always a
single BF16 A100.

\subsection{End-to-end Performance}
\label{sec:eval-compare}
\autoref{fig:goodput_4way} compares all three systems across three metrics
and four model scales: SLO attainment vs.\ offered rate, attainment
as the joint TTFT/TPOT SLO is scaled from loose to strict, and TTFT/TPOT
latency at the reference rate. HMA-Serve delivers the
highest goodput at 8B ($2.3\times$ over the strongest baseline), 14B ($3.2\times$),
and 32B ($1.85\times$ over homogeneous disaggregation and $1.1\times$ over the
oracle colocation), because as the prompt lengthens the lone prefill A100 saturates
while the parallel GDDR mesh sustains a far higher rate before TTFT breaches the
SLO. The one regime HMA-Serve does not win on raw goodput is the smallest 4B
model: a single A100 prefills a 512-token prompt in tens of milliseconds---faster
than even four parallel GDDR chips---so homogeneous disaggregation leads
($24$ vs.\ $45.6$~req/s, under a bar set to $5\times$ its \emph{own} no-load
latency). This is purely a prefill effect, not a decode limit: data parallelism
already turns a rout into a $\sim\!2\times$ gap, and the cost analysis below nearly
erases it. The crossover is governed by prefill size: once the prefill saturates a
single A100 (8B and above), cheap parallel GDDR prefill wins decisively; the
homogeneous baselines also reproduce the Sarathi-Serve finding that chunked-prefill
colocation matches or beats 1P1D disaggregation up to 14B, and only in the most
prefill-heavy 32B regime---where colocation's prefill chunks inflate per-token
latency under a tight interactive TPOT bound---does phase isolation let
disaggregation overtake it.

\subsection{Cost Efficiency}
\label{sec:eval-cost}
\begin{figure}[t]
    \centering
    \includegraphics[width=\linewidth]{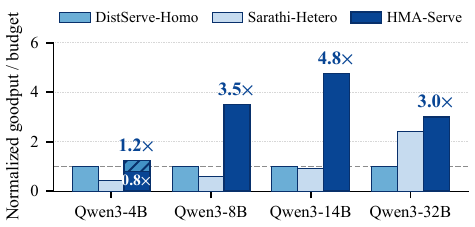}
    \caption{\textbf{Goodput per hardware budget}, each system's gp@90 divided by
    its cost in A100-equivalents (one A100 $=1.0$, one p150 $=1/12$). HMA-Serve
    and the oracle colocation share the same heterogeneous box (one A100 + the
    four-chip mesh, cost $1.33$); DistServe-Homo pays for a second A100 (cost
    $2.0$). Bars are normalized to DistServe-Homo ($=1$), so each HMA-Serve bar's
    height is its cost-efficiency advantage. The hatched 4B cap is a projection to
    an eight-chip data-parallel prefill configuration (zero cross-chip
    interference measured at four chips).}
    \label{fig:cost}
\end{figure}
\autoref{fig:cost} recasts goodput as goodput per hardware dollar, charging each
accelerator by its cost: an A100 is $1.0$ and a Tenstorrent p150 is $1/12$ of one,
so HMA-Serve's one-A100-plus-four-p150 box costs $1.33$ A100-equivalents against
the two A100s ($2.0$) homogeneous 1P1D requires. Under this accounting HMA-Serve's
advantage widens: by retiring the second A100 in favor of cheap prefill silicon it
serves $3.5\times$, $4.8\times$, and $3.0\times$ the goodput-per-dollar of
DistServe-Homo at 8B, 14B, and 32B, and $1.2$--$5.7\times$ that of the oracle
colocation---which runs on the \emph{same} $1.33$-cost box yet extracts far less.
Only 4B still favors homogeneous hardware on measured cost ($0.8\times$), but the
gap is far narrower than the raw-goodput gap because the budget saved by replacing
the prefill A100 with GDDR offsets most of the A100's prefill edge---and it closes
under scale-out: each added p150 contributes only $1/12$ of an A100 while prefill
throughput scales linearly, so growing the data-parallel prefill from four to eight
chips is projected to lift 4B to $1.2\times$ over DistServe-Homo even at the
smallest scale.

\subsection{Accuracy and Quality}
\label{sec:eval-accuracy}
\autoref{tab:quality} compares generation quality on three reasoning benchmarks
(MATH500 and AIME24/25) at three precisions: full \textbf{BF16}; a fully quantized
\textbf{BFP8} pipeline (both phases in BFP8, as a Tenstorrent-native deployment
would run); and \textbf{HMA-Serve} (BFP8 prefill, BF16 decode via deferred
dequantization). HMA-Serve tracks full BF16 across all three benchmarks on both
Qwen3-32B and 8B, while fully quantizing both phases (BFP8) costs accuracy on the
harder AIME problems---confirming that keeping the error-accumulating decode phase
in BF16 preserves quality while still reaping the cost and bandwidth benefits of
low-precision prefill, and realizing across vendors the same phase-asymmetric
precision split that Mix-Quant~\cite{mixquant} demonstrates on a single GPU.

\begin{table}[t]
\centering
\footnotesize
\begin{tabularx}{\linewidth}{@{}ll *{3}{>{\centering\arraybackslash}X}@{}}
\toprule
Model & Precision & MATH500 & AIME24 & AIME25 \\
\midrule
\multirow{3}{*}{Qwen3-32B}
 & BF16      & \textbf{97.4} & \underline{85.3} & \textbf{86.7} \\
 & BFP8      & \underline{97.3} & 83.3 & 84.7 \\
 & HMA-Serve & \textbf{97.4} & \textbf{86.7} & \underline{85.3} \\
\midrule
\multirow{3}{*}{Qwen3-8B}
 & BF16      & 93.7 & 75.5 & \underline{67.8} \\
 & BFP8      & \underline{94.1} & \underline{76.5} & \textbf{75.3} \\
 & HMA-Serve & \textbf{94.4} & \textbf{76.7} & 66.7 \\
\bottomrule
\end{tabularx}
\caption{\textbf{HMA-Serve preserves model quality} on three reasoning benchmarks.
Per (model, benchmark) the best is in \textbf{bold}, second \underline{underlined};
HMA-Serve (BFP8 prefill, BF16 decode) tracks full BF16, while fully-quantized BFP8
costs accuracy on the harder AIME problems.}
\label{tab:quality}
\end{table}

\section{Related Work}
\textbf{Single-instance batching.} Early serving systems run both phases on one
instance and maximize batching. Orca~\cite{orca} introduced iteration-level
continuous batching, and Sarathi-Serve~\cite{sarathi} added chunked prefill to fold
partial prefill into ongoing decode batches; both achieve high throughput but
produce variable latency, as each prefill blocks co-located decode steps.

\noindent \textbf{Prefill--decode disaggregation.} Observing that the phases have opposing
latency and throughput needs, DistServe~\cite{distserve} and Mooncake~\cite{mooncake}
run them on separate instances linked by a KV-cache transfer, eliminating
interference at modest cost---but assume a homogeneous pool of identical HBM GPUs.

\noindent \textbf{Heterogeneous hardware.} Splitwise~\cite{splitwise} lets prefill and decode
pools use different GPU \emph{generations} (e.g.\ older H100s for prefill, newer
H200s for decode), exploiting decode's greater bandwidth sensitivity; both sides
remain HBM-based, so the HBM cost premium is paid in full.

\noindent \textbf{Phase-asymmetric precision.} Mix-Quant~\cite{mixquant} observes that the
two phases also differ in precision sensitivity, running prefill in NVFP4 and decode
in BF16 on the same NVIDIA Blackwell GPU; the underlying hardware is unchanged.

\noindent \textbf{Toward MemHA serving.} SPAD~\cite{spad} proposes dedicated phase-specialized
silicon---a GDDR prefill chip paired with an HBM decode chip---and argues this is
the most cost-efficient point, but is a simulation study with no real pairing. A concurrent proposal~\cite{multivendorpd} likewise targets prefill--decode disaggregation across GPUs from \emph{different vendors}, but, like SPAD, evaluates its design in simulation and addresses neither the cross-vendor KV-format mismatch nor a real low-precision serving path.
HMA-Serve is, to our knowledge, the first to demonstrate cross-vendor MemHA serving
on real silicon, and the first to make the precision asymmetry between vendors a
performance lever rather than a compatibility obstacle.

\section{Conclusion}
We presented HMA-Serve, a memory-heterogeneous serving system that
pairs cheap GDDR-based Tenstorrent accelerators for prefill with an HBM-based NVIDIA
GPU for decode. By co-designing phase-wise quantization, a layer-wise
compute-transfer pipeline, and deferred dequantization, HMA-Serve hides the
cross-vendor KV path and turns the precision asymmetry into a performance lever,
achieving up to $3.2\times$ higher goodput and $4.8\times$ higher goodput-per-dollar
than homogeneous A100 disaggregation with no measurable quality loss. More broadly,
HMA-Serve shows that the most cost-efficient point in the LLM-serving design space
lies \emph{across}, not within, a single vendor's stack, and that the precision and
layout mismatches this entails can be absorbed entirely off the critical path.

\bibliography{refs}

\end{document}